\documentclass{scrartcl}
\usepackage{graphicx}        
\usepackage{natbib}         
\usepackage{color}           
\usepackage{url}             

\usepackage{ulem}
\begin{document}


\begin{center}

\textbf{The Influence of Solar Flares on the Lower Solar Atmosphere: Evidence from the Na D Absorption Line Measured by GOLF/SOHO  }\newline

G.~Cessateur$^{1}$, M.~Kretzschmar$^{1}$, T.~Dudok~de~Wit$^{1}$, P.~Boumier$^{2}$\newline
\end{center}

$^{1}$ LCP2E  (UMR 6115), CNRS and University of Orl\'eans, 3A avenue de la Recherche Scientifique, 45071 Orl\'eans cedex 2, France, 
                      email:\url{gael.cessateur@cnrs-orleans.fr}\newline
$^{2}$ Institut d'Astrophysique Spatiale (IAS), (UMR 8617), CNRS and Université Paris-Sud 11, Batiment 121, F-91405 Orsay, France

\begin{abstract}

Solar flares presumably have an impact on the deepest layers of the solar atmosphere and yet the observational evidence for such an impact is scarce. Using ten years of measurements of the Na D$_{1}$ and Na D$_2$ Fraunhofer lines, measured by GOLF onboard SOHO, we show that this photospheric line is indeed affected by flares. The effect of individual flares is hidden by solar oscillations, but a statistical analysis based on conditional averaging reveals a clear signature. Although GOLF can only probe one single wavelength at a time, we show that both wings of the Na line can nevertheless be compared. The varying line asymmetry can be interpreted as an upward plasma motion from the lower solar atmosphere during the peak of the flare, followed by a downward motion.
\end{abstract}

\section{Introduction}\label{sec:intro}

Solar flares are the most powerful events occurring in the solar system. The mechanism invoked for their energy release is the conversion of magnetic energy into radiation, thermal and kinetic energy through the reconnection of magnetic field lines. More precisely, the coronal magnetic field suddenly changes its configuration and free energy when the stress imposed at its footpoints by photospheric plasma motion exceeds a certain threshold \citep{2005ApJ...635.1299A}. The details of why and how the energy is released are still largely debated (\textit{e.g} \citep{2009AdSpR..43..739S}). Once converted, the kinetic energy accelerates particles; part of these are directed downwards and deposit their energy in the solar atmosphere while another part can eventually escape into interplanetary space. The interaction of these accelerated particles with the surrounding plasma ultimately results in electromagnetic radiation at various wavelengths, from X-ray to radio domains \citep{2008LRSP....5....1B}.

One of the best observed manifestations of flares is the appearance, after the impulsive phase, of loops that emit strongly in the extreme ultraviolet and soft X-rays (SXR). Downward-accelerated particles lose their energy by collisions when they encounter the dense chromospheric plasma. The local material is heated, expands, and then rises to fill the newly configured magnetic-field loop where plasma cools down through radiative and conductive losses. This is the model of chromospheric evaporation \citep{1984ApJ...287..917A,1987ApJ...317..502F}: the hot plasma located just above the energy-deposition layer evaporates towards the corona. The cool plasma just underneath is pushed down; this is called chromospheric condensation. In this model, the energy deposition layer is located in the middle chromosphere or in the transition region. The rise of chromospheric plasma is manisfested by the observation of Doppler-shifted spectral lines that form at these temperatures \citep{1987ApJ...317..956S,2005A&A...430..679B} 
and by the downflow of plasma as observed from the H$\alpha$ line \citep{1987SoPh..113..307F}.\\

White-light (WL) continuum emission has been also observed in some flares -- that are then called white light flares (WLF) -- but their origin is unclear \citep{2006SoPh..234...79H}. It is generally agreed that the WL emission takes place in the lower atmosphere, near the minimum temperature region or below \citep{1986A&A...156...73A,2003A&A...403.1151D}. How these layers are heated, however, remains unknown (see the review by \citep{2007ASPC..368..417D}). Among the various proposed mechanisms, direct heating by non-thermal electron beams, chromospheric radiative back-warming, or H$^{-}$ continuum emission \citep{2006ApJ...641.1210X,2000A&A...354..691G} are often cited. When taken individually, none of these mechanisms can provide the amount of energy required by the observations. Direct heating in or near the photosphere has also been suggested \citep{1994ApJ...429..890D,2009AdSpR..43..739S}. The latter mechanisms suggest that the low solar atmosphere (the chromosphere of course, but also the temperature minimun region and the upper photosphere below) is heated during these flares. As technology evolves, more WL emission is observed during flares \citep{2008ApJ...688L.119J,Kretzschmar_sub}. 
We present here observations in visible light made by the \textit{Global Oscillations at Low Frequency} (GOLF:~Gabriel \textit{et al}, 1995) instrument onboard SOHO (space mission of international cooperation between ESA and NASA) since 1996 \citep{1997SoPh..175..207G}. These observations give information on the low solar atmosphere during solar flares.\\

One question that we address in this paper is the recurrent problem of retrieving a global picture of the flare since, in addition to the intrinsic variability of the events, the conditions (position, amplitude, ...) vary and the observations
are often made by different instruments. We follow here the same approach as in Kretzschmar \textit{et al}, (2010) and search for a statistical signature of flares by analysing ten years of SOHO/GOLF data consisting in the intensity of the Fraunhofer absorption 
Na D lines integrated over the full solar disk. These lines are formed in the solar photosphere below the temperature minimum region \citep{1980A&A....92...63C,1992A&A...265..237B}. We refer here to the photosphere as a few hundred kilometer thick region between the altitude where the optical depth at 500 nm is unity ($ \tau_{500} = 1 $) and the temperature minimum region.  More precisely, the wings of the Na D line that we study in this paper are mostly generated between 300 and 500 km, just below the transition region and the 
chromosphere \citep{2001A&A...371.1128E}. Simulations and observations for a quiet Sun atmosphere have been recently made for the formation of the Na D lines \citep{2010ApJ...709.1362L}: this study strongly contributes to the idea that these lines are originating from the photosphere. However during a flare, the formation heights of the Na D sodium may differ from quiet conditions. Nevertheless changes in the Na D lines during flares thus indicate that flares affect the lower solar atmosphere including both photosphere and chromosphere. Our analysis therefore provides an opportunity to study the influence of solar flares on the lower
solar atmosphere. This is particularly interesting since the best observed flare features usually occur in the upper solar atmosphere, where the contrast between the flare signal and the background emission is largest.\\

The intensity of the Sodium line has been measured every five seconds by GOLF. The good cadence, duration and duty cycle of the observations, together with the high quality associated with space-based observations, partly allow us to compensate for the lack of spatial resolution. As we shall see below, large and even medium amplitude flares do have a significant impact on the Na line, which we interpret as the signature of photospheric heating. Furthermore, this work also points to the possible influence of flares on helioseismic studies. \\

In Section 2 we introduce the GOLF instrument, the data, and the analysis technique. In Section 3, the results are presented and discussed. Conclusions follow in Section 4.

\section{Instrumentation and Data Analysis}

\subsection{The GOLF Instrument}
GOLF, which is one of the three helioseismic instruments onboard SOHO, is a resonant-scattering spectrophotometer that monitors periodic variations of the solar surface by observing the Doppler shift of the Na D$_1$ and Na D$_2$ Fraunhofer lines respectively at 589.6 and 589 nm integrated over the solar disk. The GOLF instrument cannot distinguish these two lines and so it monitors the sum of the signals coming from both lines. These two lines have approximately the same behaviour, and so we can have an idea of the evolution of one absorption line even when working with the sum. The principle of this device, however, is best understood by considering one single line. 

\begin{figure}[!ht]
\centerline{
\noindent \includegraphics[width=7cm,height=6cm]{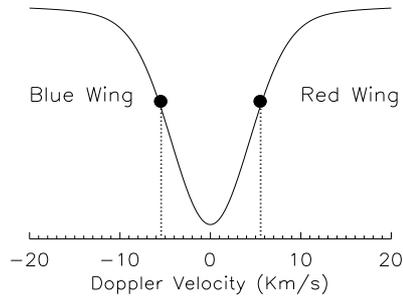}}
\caption{Location of the two wavelengths measured by GOLF. When working with Doppler shifts, one commonly expresses wavelengths in terms of Doppler velocities. The measured intensity corresponds to $\pm$ 5.5 km s$^{-1}$ around the wavelength at rest.}
\label{representation_line}

\end{figure}

GOLF was designed to measure the solar flux in the wings of the absorption line, at $\pm$ 0.0108 nm from the central rest wavelength as shown on Figure~\ref{representation_line}. By comparing the intensity of the two wings, the Doppler shift can be retrieved and the frequency of solar surface oscillations can be determined.

GOLF initially measured alternatively the intensity of each wing every 20 seconds. Unfortunately, due to a malfunction of the device, a single-wing working cycle had to be adopted to ensure the safety of the instrument. We consider here red-wing observations from 15th, October 1998 to 17th, November 2002 and blue-wing observations from 11th, April 1996 to 28th, June 1998 and from 18th, November 2002 to 23rd, May 2006. Note that an observed blue shift (negative velocity by convention) corresponds to a motion towards the observer, \textit{i.e.} upward in the solar atmosphere. In the following, the SXR flux is measured by the SEM instrument onboard GOES in the 0.1 -- 0.8 nm band.
 
\subsection{The Method: Conditional Averaging}

The signature of a single flare has no discernible effect on the intensity measured by GOLF because it is masked by omnipresent temporal fluctuations at the photospheric level, which are caused by internal acoustic waves and by granulation. GOLF was designed to study these pressure waves, but for us they are background noise that mask the flare signal.
To overcome this problem, we refrain from looking at individual events and consider the flare signature in a statistical sense. We use a conditional-averaging technique in order to see the effects of solar flares. This technique has frequently been used to detect coherent structures in turbulent flows \citep{2006PhST..122...25B} and is called superposed epoch analysis in the frame of geophysics. By averaging $N$ time sequences that are synchronised with respect to the same trigger (in our case the occurrence of the flare, as evidenced by the peak intensity of SXR flux), incoherent fluctuations are averaged out whereas reproducible features such as flare signatures add up.

We sorted the flares in decreasing order of amplitude, from the most powerful X-class flares down to the weaker M, C, and B-class flares, by using the soft X-ray classes derived from GOES observations. The largest flares are often accompanied by solar energetic-particle events that perturb the instruments. To avoid a possible impact on our results, we discarded flares whose peak energy exceeded $ 5 \times 10^{-4}$ Watts m$^{-2}$ and checked that the remaining flares were not accompanied by such particle events. We excluded in that way the ten largest flares out of a total number of 116 X-class flares. Table \ref{tab:sample} indicates the number of events observed in each wing. \\

\begin{center}
\begin{table}
\caption{ Number of events for different class-flare and the wing in which it is observed
}
\label{tab:sample}
\begin{tabular}{ccclc}     
  \hline                   
 &X & M & C & B  \\

  \hline
	Red wing & 54 & 917  & 8585 & 1013 \\
  Blue wing & 52 & 467  & 3804 & 5871 \\ 
  \hline
\end{tabular}
\end{table}
\end{center}

\section{Results and Discussion}

\subsection{Observed Changes in the Intensity}

\begin{figure}[!ht]
\centerline{
\includegraphics[width=0.6\textheight,height=0.5\textwidth,clip=]{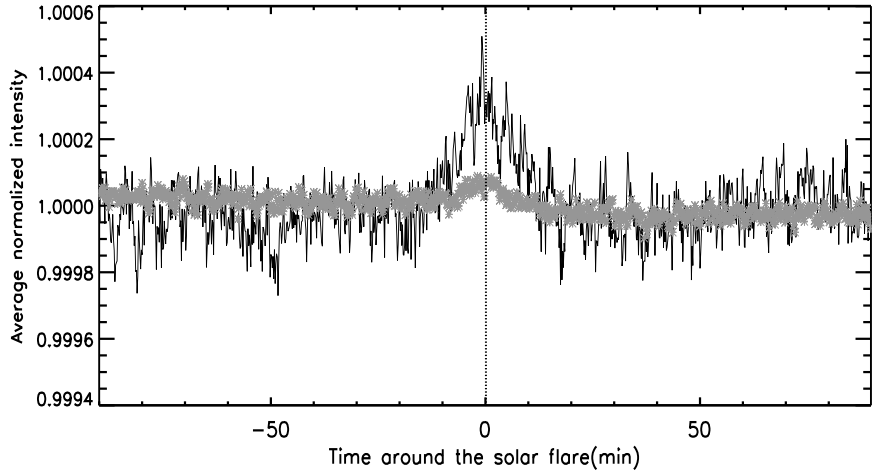}}
\caption{Conditionally averaged intensity of the red wing, measured for X-class flares (dark thin line) and M-class flares (grey line)}
\label{Results_method_X_M}

\end{figure}

The results of the conditional averaging are summarised in Figure~\ref{Results_method_X_M}, which shows the average intensity of the red wing measured for 54 X-class flares (in black), and for 917 M-class flares (in gray). Time zero [$t=0$] corresponds to the peak intensity of the SXR flux. For each flare occurring at time $t_f$, we isolated a sequence running from $t_f - 90$ minutes to $t_f+90$ minutes and averaged it.

Figure~\ref{Results_method_X_M} reveals a small but significant increase in the intensity, which cannot be observed in any individual event because it is masked by fluctuations. The relative increase in intensity is about 0.03\% for X-class flares and 0.008\% for M-class flares. Similar results are obtained for the blue wing. We conclude that flares do have an influence on the sodium line and thus on the lower atmosphere. C-class flares also have a discernible effect on the intensity of the sodium line but their influence is too weak to be properly studied and so we shall not consider them further.

GOLF measures the intensity of both wings of the line, integrated over the full solar disk. As the solar flare occurs over a small area on the Sun (less than 1\% of the total surface), the resulting intensity is the sum over the intensity originating from in the quiet Sun, which is by far the most important contribution, and the intensity originating from the flaring area. This explains why the increase in intensity is relatively weak.

\subsection{What Causes the Intensity to Change?}

Several mechanisms can be invoked to explain the observed intensity change during flares. The main ones are: 

\begin{itemize}
\item The temperature minimum region and the upper photosphere are heated, which results in an excess continuum emission. This emission can originate from Paschen and/or H$^{-}$ continuum. This effect has been invoked to explain WL emissions during solar flares \citep{2006ApJ...641.1210X,2006SoPh..234...79H}. We then expect the entire intensity profile of the sodium line to increase and so both wings should vary by the same amount.

\item Sodium is heated and excited or ionised. This causes a depletion of the absorbing sodium and hence an increase in the intensity of the absorption line. The resulting signature is the same as before. 

Hydrogen lines are known to change sometimes from absorption to emission during very large X-class flares \citep{1976sofl.book.....S}. These lines, however, are mostly generated in the upper chromosphere where more energy is released than in the upper photosphere. As the Na D lines are generated in the upper photosphere, we expect them to remain in absorption for all the flares that we consider.

\item Sodium is heated and some material is rising. This causes both a Doppler shift and an intensity change. The upward flow of sodium shifts the line towards the blue and so the intensity decreases in the blue wing with respect to the red one.
\end{itemize}

Other effects may contribute to this. The recombination of chromospheric sodium ions, for example, which is enhanced by the flare-associated density increase, may shift the sodium line to the blue and hence lead to a decrease in the blue wing. 

If the GOLF sodium line had been measured simultaneously at different wavelengths (as will be the case with the proposed GOLF-NG instrument \citep{2006AdSpR..38.1812T}), then a partial separation between these different mechanisms will be possible. Our observations, which are based on the measurement of one single wavelength at a time, \textit{a priori} totally exclude such a separation.  The important point, however, is that even though we do not measure the red and blue wings simultaneously, we do observe their average temporal evolution under similar conditions. For that reason, the two wings are directly comparable in a statistical sense even though individual events are never recorded simultaneously. A few assumptions, however, need to be made before we can pursue with such a comparison.
 
First, our events occur at different times, so we have to  assume that the impact of flares on the photosphere is not solar-cycle dependent. We can indeed reasonably assume that the speed of plasma motions should be about the same, regardless of the phase with the solar cycle. 

Secondly, for our samples to be statistically comparable, the number of events and the average SXR flare intensity should be about the same in both categories. We checked this and the sizes of the samples, which are given in Table~\ref{tab:sample}, are indeed sufficiently close. In particular, the noise level, which is proportional to $1/\sqrt{N}$ (where $N$ is the sample size) should be approximately equal in both wings, for a given class of flares.

Since GOLF observes the full Sun, the sodium line integrates contributions coming both from the quiet Sun and from active regions. Because of the statistical procedure, however, the former are averaged out whereas the latter add up coherently. Obviously, the observed variation in the integrated intensity is considerably smaller than that observed in the vicinity of the flare.

Another point is the location of the solar flare on the Sun. Because of projection effects, flares occurring at the center of the disk and at the limb have different Doppler shifts along the line of sight. The shift associated with radial motion is largest for flares located close to the center of the disk. To avoid such geometric effects, we restricted our sample to flares whose heliocentric angle is less than 30 degrees.

\subsection{Interpretation of the Intensity Increase}

Let us start with X-class flares, whose signature on the sodium line is most evident. \\

\begin{figure}[!ht]
\centerline{
 \includegraphics[width=1\textwidth,clip=]{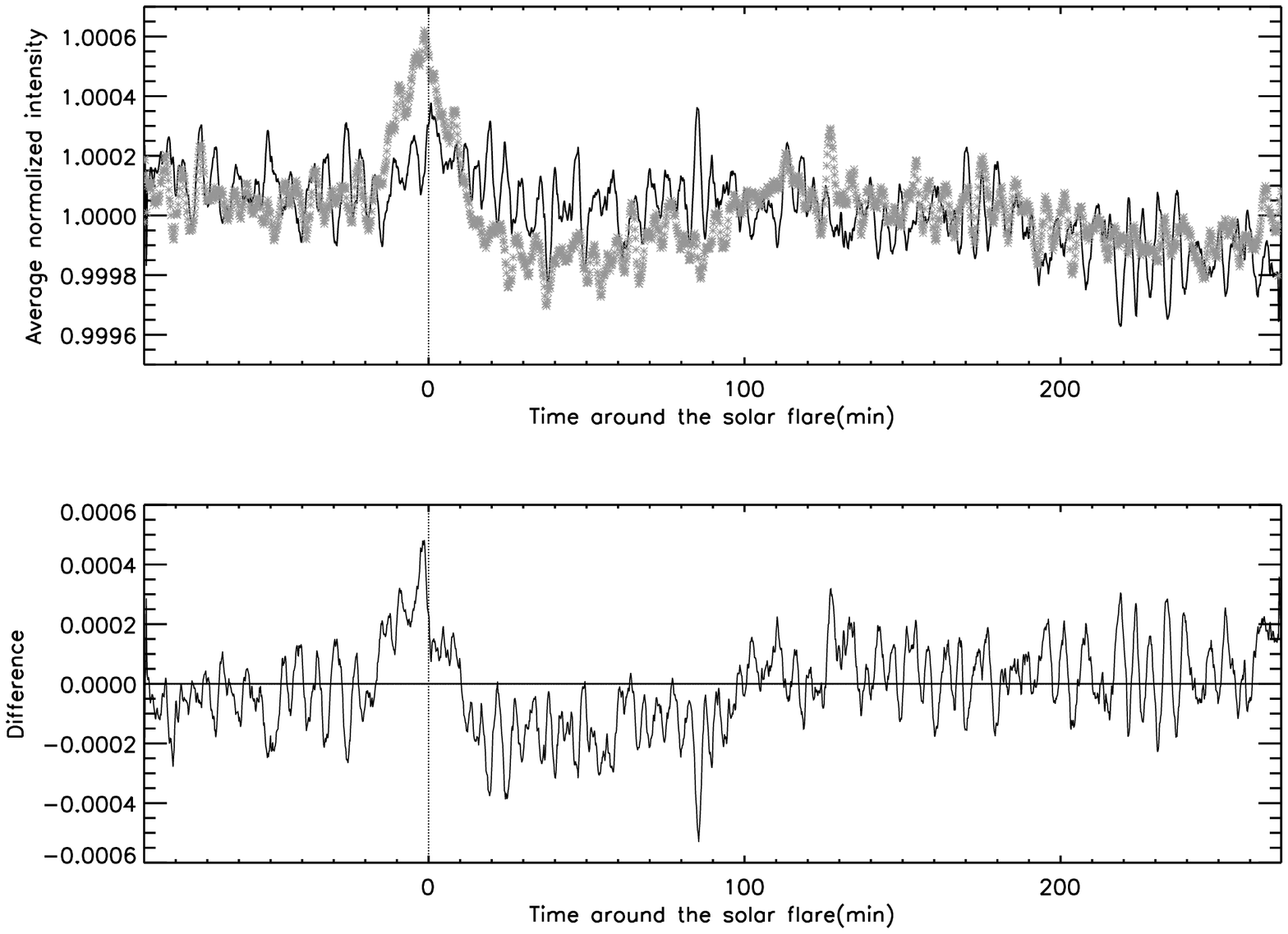}}
\caption{Top panel: conditionally averaged intensity of the red (in gray) and the blue (in black) wing for the X-class flares. Curves have been smoothed over two minutes. Bottom panel: intensity difference between the two wings (I{\tiny red} - I{\tiny blue}) }

\label{comparison_X}
\end{figure}

Figure~\ref{comparison_X} shows the conditionally averaged intensities of both wings, together with their difference (\textit{e.g} I{\tiny red} - I{\tiny blue}). The two samples have similar sizes (see Table~\ref{tab:sample}) and similar average X-rays fluxes, so a statistical comparison is meaningful. The standard deviations for both wings are very close. Moreover, the slow degradation of GOLF implies that instrumental errors should be about the same in both samples. \\

\begin{figure}[!ht]
\centering{
\includegraphics[width=0.32\textwidth]{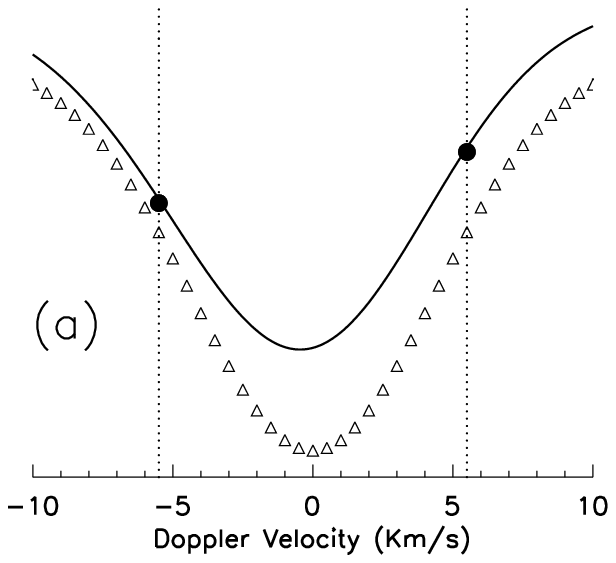}
\includegraphics[width=0.32\textwidth]{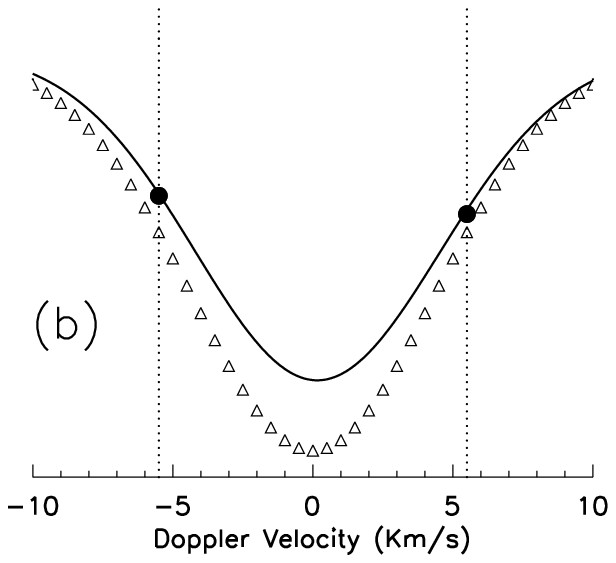}
\includegraphics[width=0.32\textwidth]{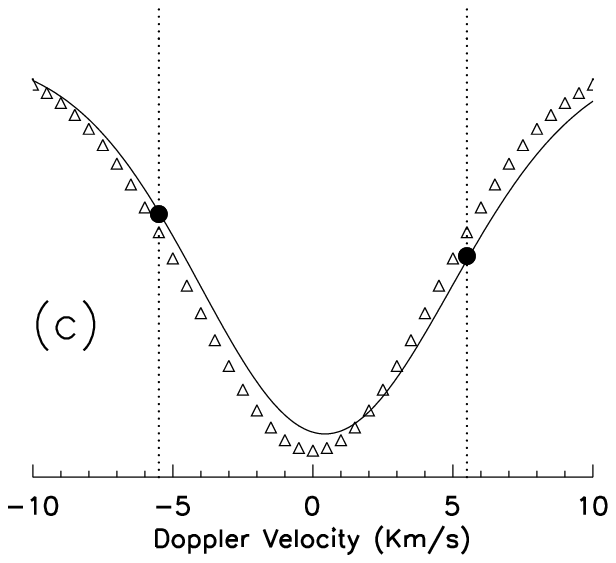}
}
\caption{Schematic representation of the possible causes for the intensity changes observed in Figure 3. The triangles represent the absorption profile at rest. (a) blueshift: Upward motion to chromosphere. (b) redshift: Plasma is falling. (c) Reduced excitation of sodium: whole profile rises down, always with a redshift.}
\label{comparison}
\end{figure}

Interestingly, the blue and red wings evolve differently in time. The intensity of the red wing is clearly higher at the flare maximum as compared to the blue wing. After the maximum, the reverse occurs, and the intensity of the red wing falls off more rapidly. Our interpretation is sketched in Figure~\ref{comparison}. The line represented in this sketch can be either Na D$_1$ or Na D$_2$, as we assume that they behave similarly.\\

First, the intensity of the whole line profile rises, due to the excitation and heating of the sodium. The intensity of both wings indeed increases during a flare. For that reason, Figures 4a, 4b, and 4c present a higher profile than for the quiet Sun. Moreover, the intensity difference at the peak of the flare suggests that the sodium profile is asymmetric. A blueshift occurs, which means that sodium is rising, see Figure~4a. After the peak of the flare, a deficit in intensity occurs in the red wing, which implies a downward motion of the plasma. A redshift may also occur, as represented in Figure~4b, meaning that the plasma is falling. Finally, the excitation diminishes, see Figure~4c.
Note that the lack of observation of the core of the line prevents us from concluding about global changes in the line profile. The pertinent result, however, is the difference between both wings, which can be unambiguously attributed to a Doppler shift.\\

The intensity of both wings comes back to pre-flare values in approximatively 100 minutes, as shown in Figure \ref{comparison_X}. Not surprisingly, this duration is close to that observed in the SXR flux. Figure \ref{comparison} is meant to be schematic only. We do not know, for example, what is really happening in the core of the line.\\

 \begin{figure}[!ht]
\centering{
\includegraphics[width=1\textwidth,clip=]{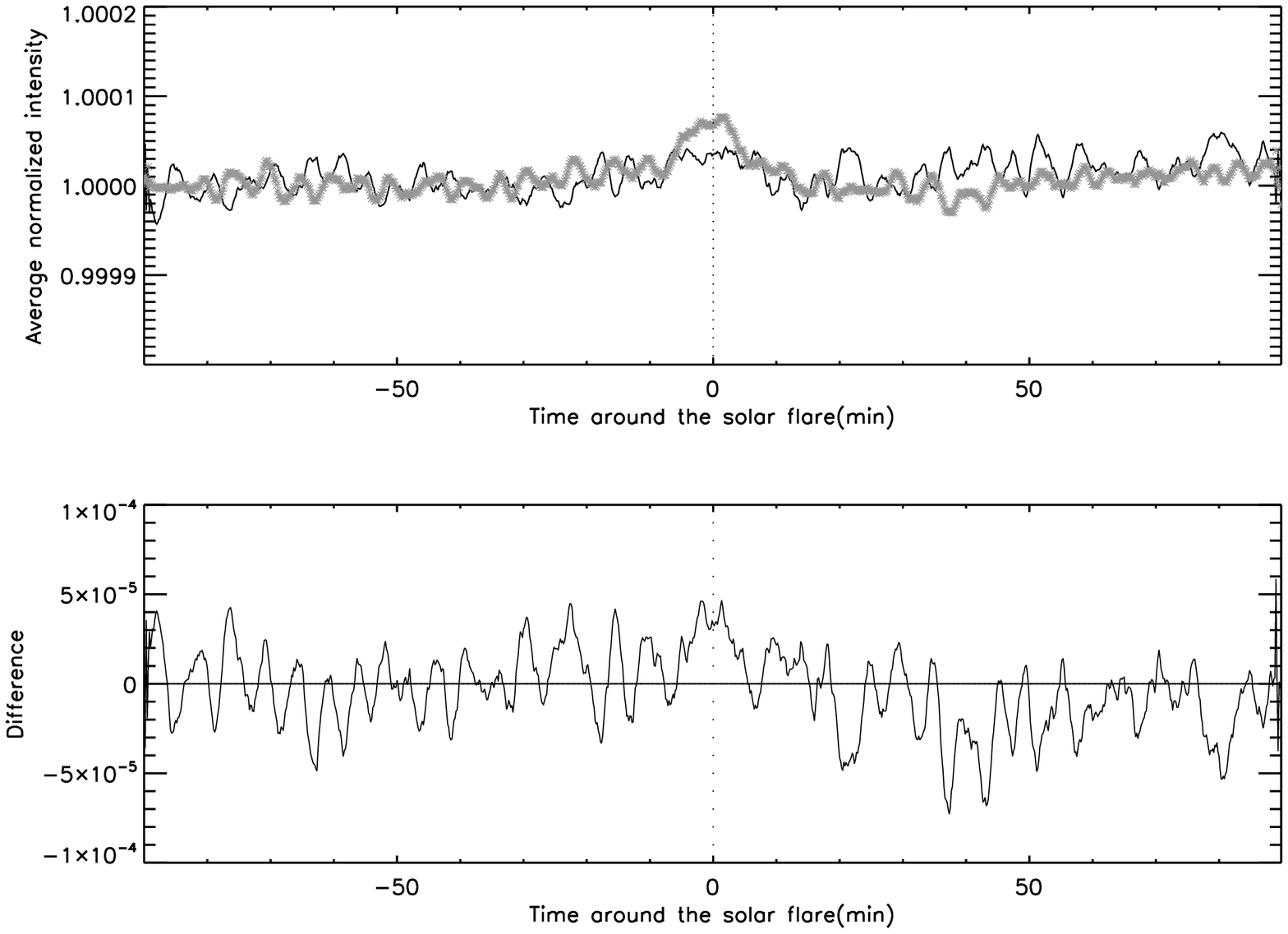}}
\caption{Top panel: conditionally averaged intensity of the red (in gray) and the blue (in black) wing for the M-class flares. Curves have been smoothed over two minutes. Bottom panel: intensity difference between the two wings.
}
\label{comparison_m} 
\end{figure}

 Similar conclusions apply to M-class flares (Figure \ref{comparison_m}), with a lower signal-to-noise ratio and a shorter recovery time. The difference between the red and the blue wings (bottom panel) is not properly resolved during the impulsive phase of the solar flare, while the redshift
 is clearly visible. These results provide direct evidence for the impact of flares on the photosphere and not just on the chromosphere. \\

Photospheric sodium is affected by several effects during a solar flare. All of these effects (excitation, heating, motion) can change the profile of the Na D lines. The speed of ascent could be determined using the Doppler shift caused by the plasma motion. 
These effects however are simultaneous, which means that the change of intensity is not just due to the Doppler shift. Furthermore, we sample the line at two wavelengths only, which is not enough to distinguish, if they were simultaneous the contribution from heating from that of the Doppler shift. \\

Our interpretation of the plasma motion, with first an upflow and then a downflow of the sodium during the flare, in based on the hypothesis that the Na D lines remain in absorption. We already discarded flares whose peak energy exceeds $5 \times 10^{-4}$ Watts m$^{-2}$ in order 
to exclude the most energetic events. Since the behaviour of the lines seems to be the same for X-class and M-class flares, one can reasonably assume that the Na D lines always remain in absorption for the events under consideration.

These results are \textit{a priori} in contradiction with the observations of downflow in the H$\alpha$ line \citep{1987SoPh..113..307F}. Indeed, in the classical picture of chromospheric evaporation, the energy deposition layer is located in the middle of the chromosphere, and so one would expect 
photospheric lines such as Na D to behave similarly to the H$\alpha$ line.
A reconciliation between these two pictures is difficult unless:

\begin{itemize}

\item {The sodium affected by solar flares is not the photospheric sodium normally monitored by GOLF for a quiet Sun, but the recombined chromospheric sodium that emits during solar flares. The lines observed by GOLF should then behave like a chromospheric one.
The disagreement with the H$\alpha$ lines then is not so surprising}

\item {The affected sodium originates from the photosphere. Our results then suggest that at least part of the energy can reach lower altitudes than expected in the classical picture. The upper photosphere can indeed be heated by several 
mechanisms (\citep{1986A&A...168..301A}, \citep{2007ASPC..368..417D}) leading to white light emission from both the chromosphere and the photosphere. Recent studies strongly suggest that these white-light flares are more
common than expected \citep{2007ApJ...656.1187F,2008ApJ...688L.119J}. As we do indeed analyze the average over several flares, and not a particular event as for the H$\alpha$ line, white-light flares could have a predominant weight in our statistics. }

\end{itemize}

These results and their interpretations are based on only two points in the spectral profile, so the conclusion on the sodium dynamics in the lower solar atmosphere needs to be confirmed by further studies. Our results nevertheless strongly suggest that the upper photosphere 
is affected by solar flares.

\section{Conclusions}

The aim of this study was to investigate whether solar flares impact the Na D$_1$ and Na D$_2$ Fraunhofer lines, as measured by GOLF. A statistical analysis based on conditional averaging reveals intensity increases during X and M-class flares, which demonstrates that the Na D lines
are affected. Although GOLF observes only one wing at a time, the average evolution of the red and blue wings can be compared in a statistical sense. The observed intensity differences suggest the existence of plasma motions. The blueshift that coincides 
 with the peak of the flare can been interpreted as a upward plasma motion, which is most likely associated with photospheric Na rising towards the chromosphere. The peak of the flare is followed by a redshift, which corresponds to a downward flow.
 Since the Na D line is mostly generated in the upper photosphere, our study thereby provides strong support for the influence of flares on the lower solar atmosphere.

This interpretation is \textit{a priori} in contradiction with the classical picture of chromospheric evaporation. It can however be explained if we assume that white-light flares, for which the lower solar atmosphere needs to be heated, have a preponderant effect in our statistics.
White-light flares could indeed be more common than expected, and our study supports this idea.
However, further quantitative studies are necessary before we can conclude that solar flares do really affect the lower solar atmosphere. Plasma motions could be more accurately determined with
more than two measurements on the profile of Na D lines. The GOLF-NG has been proposed for that purpose. Another important conclusion is that future helioseismic instruments should take into account the impact of solar flares in order to improve the measurement of solar oscillations.\\

\textbf{Acknowledgments}This study received funding from the European Community's Seventh Framework Programme (FP7/2007-2013)
under the grant agreement nr. 218816 (SOTERIA project, \url{www.soteria-space.eu}). The GOLF experiment is based on a consortium of institutes involving a large number of scientists and engineers,
belonging to the IAS (France), the CEA (France), the IAC (Spain), and the observatories of Bordeaux and Nice (France). SOHO is a missing of international collaboration between ESA and Nasa.

\bibliographystyle{spr-mp-sola}
\bibliography{sodium}
\end{document}